\documentclass[a4paper,fleqn,usenatbib]{mn2e}
\usepackage{amssymb}
\usepackage{textcomp}
\usepackage{graphicx}
\usepackage{txfonts}
\usepackage{epstopdf} 
\usepackage{xcolor}     

\begin{document}

\title[Period and colors of (6478) Gault]{Spinning and color properties of the active asteroid 
(6478) Gault\thanks{Based on observations collected at the Cassini Telescope of the Loiano Observatory, Italy}}
\author[Carbognani \& Buzzoni]{Albino Carbognani\thanks{E-mail: albino.carbognani@inaf.it}
\& Alberto Buzzoni\\
INAF - Osservatorio di Astrofisica e Scienza dello Spazio, Via Gobetti 93/3, 40129 Bologna, Italy
}


\date{Received ; Accepted}

\maketitle

\label{firstpage}

\begin{abstract}
We report on accurate $BVR_{c}$ observations of (6478) Gault, a 5-6 km diameter inner main-belt 
asteroid in the Phocaea family, notable for its sporadic, comet-like ejection of dust. This curious 
behavior has been mainly interpreted as reconfigurations after YORP spin-up, although merging of a contact 
binary system cannot be fully excluded. We collected optical observations along the 2019 March-April period,
at orbital phase angles between $12^{\circ}-21^{\circ}$, to search for direct evidence of asteroid quick spinning rotation.
A prevailing period value of $3.34 \pm 0.02$ hours is supported by our and other photometric observations. 
In the YORP spin-up hypothesis, this period points to a bulk density $\rho \approx 1$ $\textrm{g}/\textrm{cm}^3$. 
The mean colors are $B-V = +0.82_{\pm 0.3}$, $V-R_{c} = +0.28_{\pm 0.06}$ and $B-R_{c} = +1.11_{\pm 0.4}$, 
but we have observed a strong bluer color during the April session, with about $\Delta (B-V) \sim 0.35 \pm 0.09$~mag. 
This color change can be due to asteroid rotation and support the hypothesis that there is a bluer surface under the 
Gault's dust, as indicated by spectroscopic observations made on 2019 March 31 and April 8 by \citet{marsset2019}.
\end{abstract}

\begin{keywords}
minor planets, asteroids: individual: (6478) Gault 
\end{keywords}

\section{Introduction}
\label{sec:introduction}

\begin{figure}
\includegraphics[width=\hsize, clip=]{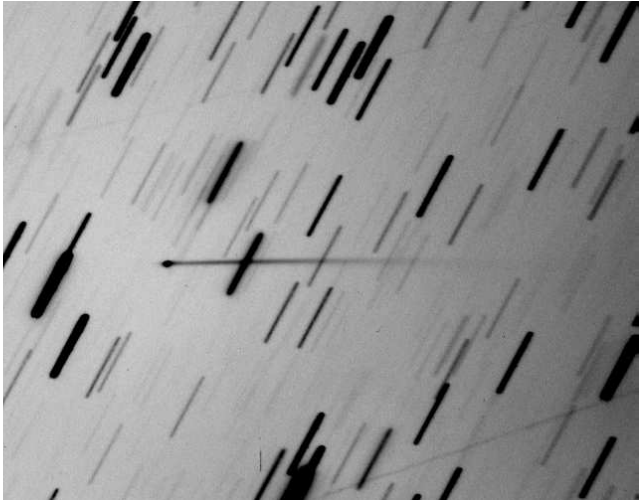}
\caption{A $14 \times 11$ arcmin picture of (6478) Gault with its tail taken from OAVdA on 2019 March 23, about 
20:20 UT ($\alpha$ = 10 h 04 m 25.2 s, $\delta$ = $-01^\circ$ $08{'}$ $06.7{''}$; J2000.0). North
is up, East to the left.
The main tail length is $5{'} 30{''}$ at position angle $PA \approx 272^\circ$. Its also visible 
a fainter $12{''}$-extended anti-tail at $PA \approx 91^\circ$. The image is a stack of 38 frames, each 
with 180~s exposure time.}
\label{f01}
\end{figure}

Main belt asteroid (6478) Gault (hereafter ``Gault''), recently surged to very special
attention \citep{smith2019} as an outstanding member of the active asteroids class, sporting 
typical morphological features of comets, such as a coma and tail, see 
Fig.~\ref{f01}. Pre-covery research in the NOAO image database \citep{chandler2019} 
allowed us to trace Gault's outbursts back to year 2013. As the outbursts appeared along the full 
heliocentric orbit, even about the aphelion distance of 2.75 AU, this feature tends to exclude 
the sublimation of volatile material as a cause of the activity. Furthermore, spectroscopic 
observations \citep{jewitt2019} showed a prevailing presence of dust, rather than gas, both in 
the coma and in the asteroid tails.

A re-iterate sequence of outbursts in the last year, namely on 2018 October $28\pm 5$, December $31\pm 5$, 
2019 February $10\pm 7$ \citep{jewitt2019}, also including part of the present observations, may  
rule out as well the unlikely case of multiple impacts 
with smaller bodies as the triggering physical mechanism of Gault's activity. Rather, this may definitely restrain
the focus to an intervening dynamical instability of the asteroid's structure, where a nearly 
spin-barrier rotation could strongly ease the on-going disintegration of a ``fluffy'' body
\citep{kleyna2019}. Alternative to any rotation-driven process, however, also binary-system merging
could be invoked as the main responsible of Gault's outbursts \citep{quanzhi2019}.
 
The presence of the spin-barrier in the ``realm of asteroids'' can be explained by the cohesionless 
``rubble-pile'' structure model, assuming asteroids to consist in fact of collisional breakup 
fragments mainly bunching together under mutual gravitation \citep{pravec2002}, but in some case perturbed
by centrifugal forces according to body's rotation speed. 
Simple physical arguments lead to estimate, for the critical rotation period ($P_S$) of a spherical 
object of bulk density $\rho$ (expressed in g\ cm$^{-3}$),
\begin{equation}
P_S = \sqrt{\frac{3\pi}{G\rho}} \approx \frac{3.3 \textrm{h}}{\sqrt{\rho}}.
\label{eq:spin_barrier}
\end{equation}

Asteroid's bulk density is a crucial but difficult parameter to obtain, as we need to know both 
mass and volume of the body. In general, S-type asteroids are denser than C-type ones, the latter likely 
displaying a larger macroporosity. Reference figures indicate
$\rho_{S} = 2.72_{\pm 0.54}$~g\ cm$^{-3}$ for S-type and 
$\rho_{C} = 1.33_{\pm 0.58}$~g\ cm$^{-3}$ for C-type objects \citep{carbognani2017}. 

Things get slightly more entangled in case of a non-spherical geometry. If we deal in particular
with the relevant case of a ``cigar-shaped'' triaxial ellipsoid (spinning around the
``$c$'' axis and with three axis constrain: $a \ge b = c$, according to \citet{richardson2005}, then the 
spin-barrier critical period ($P_E$) exceeds the spherical case of eq.~(\ref{eq:spin_barrier})
as $P_E = {\cal{F}}\ P_S$, with the shape factor ${\cal F}(\epsilon)$ fully depending on body's 
(equatorial) eccentricity \footnote{As usual, we define $\epsilon = [1-(b/a)^2]^{1/2}$, in terms of minor-to-major 
axis ratio $(b/a)$ of the body.\label{fn1}} in the form:

\begin{equation}
{\cal{F}} = \sqrt{\frac{2 \epsilon^3}{3\left( \epsilon^2-1 \right)\left( 2 \epsilon+\ln\frac{1-\epsilon}
{1+\epsilon}\right)}}.
\label{eq:spin_barrier_ell}
\end{equation}

\begin{figure}
\includegraphics[width=\hsize]{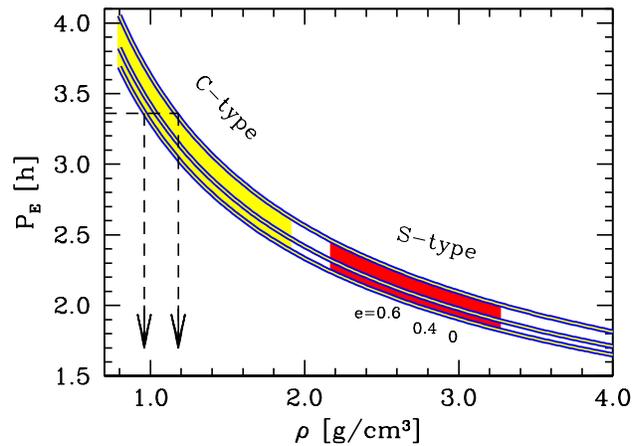}
\caption{The expected $P_E$ vs. $\rho$ relationship according to \citet{richardson2005}.
Spin-barrier critical period $P_E$ in case of a ``cigar-shaped'' triaxial ellipsoid is obtained from the spherical case
modulated by the shape factor $\cal F$ of eq.~(\ref{eq:spin_barrier_ell}), fully depending
on the body's eccentricity. In addition to the spherical geometry ($e = 0$) two cases are displayed 
in the plot, respectively with $e = 0.6$ and 0.4, with a slower critical period increasing with 
body's eccentricity, at fixed bulk density $\rho$, as labelled on the plot.
The reference bulk-density figures for C- and S-type asteroids, according to \citet{carbognani2017}
are reported as yellow and red bands, respectively. The prevailing estimate of Gault's rotation
period of $P = 3.34$~h is marked in the plot, together with the implied range for asteroid's density
(arrows) $\rho \sim 1$ g~cm$^{-3}$. See text for a discussion.
}
\label{f02}
\end{figure}

By combining eq.~(\ref{eq:spin_barrier}) and (\ref{eq:spin_barrier_ell}), a straight $P_E$ vs. $\rho$ 
relationship can be derived, as displayed in Fig.~\ref{f02} for different values
of eccentricity. According to previous bulk-density figures, one sees from the plot 
that centrifugal breakup may be reached by C-type asteroids for a spin-barrier critical period 
$P_E \sim 2.5$-4.0~h, while a shorter period, always well less than 2.5 hours, might be required for a denser  
S-type object.

No firm estimate of Gault's rotation period was available until cometary activity was first discovered
on 2019 Jan 5 \citep{smith2019}. Subsequent photometric follow up to obtain an accurate 
lightcurve of the object did not lead to any conclusive result, likely due to the masking effect
of dust in the coma \citep{kleyna2019,quanzhi2019,Man-ToHui2019,jewitt2019, sanchez2019}. 
Based on a Lomb-Scargle and ANOVA lightcurve analysis, \citet{kleyna2019} recently proposed for Gault 
a rotation period about 2 hours, which implied a density of about 2.7~g\ cm$^{-3}$ as for a typical S-type 
asteroid. However, a slower period, about 3 hours, more suitable for a C-type object, has later been claimed by \citet{ferrin2019}. 
Until now, no phased lightcurve can be reported to explicitely support any of these values.

\section{Observations \& Data reduction}
\label{sec:instruments}

Thanks to asteroid's closer distance, near opposition with Earth, and taking advantage of 
the declining trend of dust activity, we surveyed Gault along the 2019 March-April trajectory arc, with the 
purpose of determining the asteroid's rotation period from its optical lightcurve. 
A first observing batch was carried out with the OAVdA Ritchey-Chr\'etien 0.81-m f/4.75 telescope 
at Saint-Barthelemy (Aosta, Italy, MPC ID code B04) along the three nights of 2019 March 23, 26 and 27. 
The sky was with some sporadic veils the first night, while in the following two nights it was 
clear and transparent. The telescope was equipped with an FLI 1001E CCD array of $1024 \times 1024$ pixels with 24$\mu$m 
pixel size used in $2 \times 2$ binning mode such as to provide a platescale of 2.54 arcsec px$^{-1}$
across a $21.9 \times 21.9$ arcmin field of view.
Gault's imaging was performed with $C$ filter (i.e. ``white'' light), in order maximize target detection
(estimated about $V\sim 17$). The frames were dark subtracted and then flat-fielded according to the standard 
procedure. The SNR for the three sessions was near 50, the mean uncertainty are, respectively, 0.020, 0.022 and 0.023 mag.
Although fully successful ones, these observations caught the asteroid still in full
activity, with a detectable coma and an extended dust tail visible over 5.5 arcmin away at $PA\sim 272^{\circ}$, 
as well evident from Fig.~\ref{f01}.
 
\begin{figure}
\includegraphics[width=0.68\hsize, angle=-90]{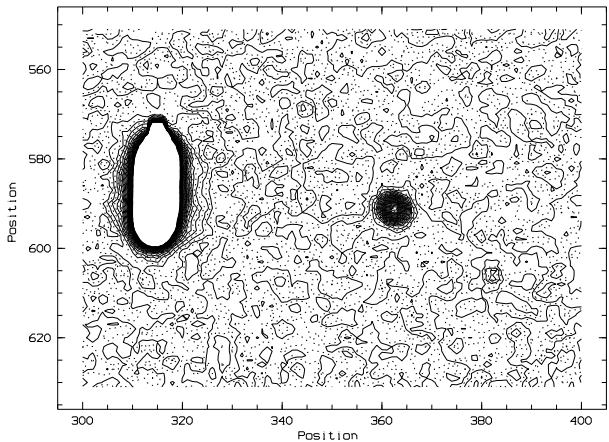}
\caption{$R_c$-band isophotal contour plot of a Gault's illustrative image from the Loiano data set,
along the night of 2019 April 15. Esposure time is 240 sec with telescope tracked at non-sidereal 
rates to follow Gault's motion. The displayed field of view is about $60\times 45$ arcsec across,
with North up and East to the left.
Coordinate axes are labelled in pixel scale (1 px = 0.58 arcsec). Gault is the ``rounded'' object 
about $(x, y) = (362, 591)$ coordinates. The vertically elongated object to the left of the
image is a saturated star distorted by on-target tracking.
Seeing on the image is about 2.2 arcsec FWHM. A bright Full Moon, only $12^{\circ}$ apart
was strongly affecting the sky background, here estimated in $\mu_R \sim 16.8$~mag~arcsec$^{-2}$. 
From the image we can however rule out at a $S/N \ge 3$ confidence level
any activity signature around the asteroid, brighter than 21.9~mag~arcsec$^{-2}$. 
}
\label{f03}
\end{figure}

A further observing run was then attempted one month later, along the night of 2019 April 15, with the 
asteroid now definitely ``turned off'' in its quiescent state (see Fig.~\ref{f03}). However, as Gault 
was becoming about one magnitude fainter with increasing its orbital phase angle, we had to rely on 
the larger ``G.B. Cassini'' 152~cm f/4.6 Ritchey-Chr\'etien telescope of the Loiano Observatory 
(Bologna, Italy, MPC ID code 598) for these new observations. The BFOSC camera was attached the 
telescope, equipped with a Princeton Instruments EEV $1340 \times 1300$ pixel back-illuminated CCD with 20~$\mu$m pixel size. 
Platescale was 0.58~arcsec~px$^{-1}$ leading to a field of view of $13.0 \times 12.6$~arcmin. 
Broad-band Johnson/Cousins $B,V,R_{c}$ filters were used to measure asteroid's colors. The telescope  
was tracked at non-sidereal rates to follow Gault's motion and increase S/N of detection.

The Loiano observations were carried out under clear but partly scattered sky, with seeing about
2.2 arcsec (FWHM) and a bright Full Moon about $12^{\circ}$ apart from the target. Nevertheless, a good sequence 
of $R_c$ images each with 240~s integration (mean uncertainty about 0.09 mag), was obtained spanning about 4~hours in total, 
interleaved by three $B,V$ series to sample asteroid's colors. The \citet{landolt1992} PG1047+003 calibration
field was taken at similar airmass of Gault images in the three $B,V,R_c$ bands, providing to avoid 
cloud interference. 
Image processing included bias subtraction and flat fielding procedure, as usual.
Due to scattered clouds, however, special care has been devoted for photometric
reduction of the entire data set, as discussed in more detail in the next sections.
Along the total of four OAVdA and Loiano observing runs we collected about 12 hours of observation 
on the target, as summarized in Table~\ref{t01}. 

\begin{table}
\centering
\caption{Summary of the 2019 OAVdA (B04) and Loiano (598) observing sessions}
\label{t01}

\begin{tabular}{lccccc}
\hline
Date & No. of & Band & Exposure & Timespan &   MPC\\
     & frames &      & [s]      & [h] & ID\\
\hline
March 23 & 47 & C & 180 & 2.0 & B04 \\
March 26 & 61 & C & 180 & 3.0 & B04 \\
March 27 & 65 & C & 180 & 3.0 & B04 \\
April 15 & 49 & $R_{c}$ & 240 & 4.0 & 598 \\
	 &  3 & $V$ & 300 &  & 598\\
         &  3 & $B$ & 480 &  & 598\\
\hline
\end{tabular}
\end{table}
\subsection{On-frame photometry}
\label{sec:phot}

MPO {\sc Canopus} package \citep{warner2009} was used for differential aperture photometry of our data.
We especially relied on the {\sc Comp Star Selector} (CSS) and {\sc DerivedMags} software feature 
to pick up a reference grid of (whenever possible) solar-type standards, from the CMC15 star 
catalog \citep{cmc15}, and therefrom lead to an accurate calibration (to within
a few hundredths of magnitude internal uncertainty) of Gault'magnitude directly on the observed field.
This is very useful because it allows the different photometric sessions to be linked together.
Gault's aperture photometry has been carried out through a $\sim 3$~FWHM circular aperture, throughout,
to account for seeing effects. Only the best frames, with the target clearly unaffected by star crowding, were retained. 
According to the CMC15/UCAC4/APASS photometric characterization \citep{carbognani2016}, we can confidently 
match the Johnson-Cousins $R_c$ system with our observations with the equation: 

\begin{equation}
R_c = r' - 0.112 - 0.128\left(B - V \right)\quad {\rm mag}.
\label{eq:rcmc15a}
\end{equation}

In eq.~(\ref{eq:rcmc15a}) $r'$ is the apparent red mag of the star in the Sloan system adopted by CMC15/UCAC4/APASS 
catalogs, while $B$ and $V$ are the mag in the Johnson system. The RMS, when using eq.~(\ref{eq:rcmc15a}), is about 0.05 mag. 
For a solar-type star, as our comparisons, $B-V \simeq 0.656 \pm 0.005$, so:

\begin{equation}
R_c \simeq r' - 0.2
\label{eq:rcmc15}
\end{equation}
 
This and the previous correction was applied throughout in the reported $R_c$ magnitudes of this paper.
A sub-set of three-to-five comparison stars across the full frame sequence for
each observing run were measured in order to assess sky transparency conditions along the
night. In particular, for the Loiano observations, this procedure allowed us to track in some 
detail the temporal behaviour of thin cloud absorption affecting Gault's imaging and recover colors to fiducialy 
cloud-free conditions. This correction is of paramount importance in order to derive the asteroid's colors 
variation.

\section{Gault's colors}
\label{sec:colors}

Three series of deeper $B,V$ images (referred to hereafter as Batch \#1, 2 and 3, with exposure time 
of 8~min in $B$ and 5~min in $V$) have been accompanying the $R_c$-band sequence along the Loiano 
session of 2019 April 15. As marked in the lower panel of Fig.~\ref{f05}, the $B,V$ luminosity was 
sampled around 20:03-20:17~UT (Batch \#1), 21:06-21:20~UT (Batch \#2), and 22:10-22:24~UT 
(Batch \#3), in order to assess Gault's apparent colors at different lightcurve phase. 
The photometric reduction has been carried out according to the usual
standard calibration procedure \citep{landolt1992, harris1981}. In addition, special care has been
devoted to take the Landolt field at similar airmass than Gault's frames
in order to minimize differential corrections.

If we look at the photometric trend of the comparison stars present in Gault's field of view we see that cloud absorption did
not affect Batch \#1, while a thinner coverage was in place, on the contrary, at Batch \#2 and Batch \#3. 
To estimate the effect of cloud absorption on colors we chose three stars, from the UCAC4 stars catalog\footnote{In the UCAC4 catalog the $B$ and $V$ mag are in the Johnson system, while the red mag are in the Sloan $r'$ system. To transform from $r'$ to $R_c$ we use eq.~(\ref{eq:rcmc15a}).}, 
placed near Gault and computed the colors with the same photometric parameters used for Gault. 
The results are shown in Table~\ref{t02a}. From this we can see how the average colors value and the one from the UCAC4 catalog are compatible within a few hundredths of magnitude. Thus, despite the presence of veils and the full Moon, the observed colors are reliable. If we look at the individual Batches, we can see how the stars colors tend to become redder, going from Batch \#1 to Batch \#3 as expected, which appears to be the most conditioned by cloud veils.
Taking as reference the Batch \#1, we can compute a set of mean correction terms defined as (colors Batch \#1)-(colors Batch \#2) or (colors Batch \#1)-(colors Batch \#3). We can use this additive terms to ``delete'' the veils effect on Gault's colors, see Table~\ref{t02b}.

\begin{table}
\centering
\caption{UCAC4 stars $BVR_{c}$ colors along the observing night of 2019 April 15. The last two columns provide, respectively, 
the average color value on three Batch and the catalog value.}
\label{t02a}
\begin{tabular}{lccccc}
\hline
473-044752 & Batch \#1 & Batch \#2 &  Batch \#3 & Average & Cat. \\
\hline
$B-V    $     & $0.33_{\pm 0.03}$ & $0.34_{\pm 0.03}$ & $0.52_{\pm 0.03}$ & $0.40_{\pm 0.06}$ & 0.43 \\
$V-R_{c}$     & $0.23_{\pm 0.02}$ & $0.27_{\pm 0.02}$ & $0.27_{\pm 0.02}$ & $0.26_{\pm 0.01}$ & 0.25 \\
$B-R_{c}$     & $0.56_{\pm 0.02}$ & $0.61_{\pm 0.02}$ & $0.79_{\pm 0.02}$ & $0.66_{\pm 0.07}$ & 0.68 \\
\hline
473-044753    &      &      &      &                   &      \\
\hline
$B-V    $     & $0.75_{\pm 0.04}$ & $0.76_{\pm 0.04}$ & $0.96_{\pm 0.04}$ & $0.82_{\pm 0.07}$ & 0.90 \\
$V-R_{c}$     & $0.39_{\pm 0.02}$ & $0.46_{\pm 0.02}$ & $0.44_{\pm 0.02}$ & $0.43_{\pm 0.02}$ & 0.41 \\
$B-R_{c}$     & $1.14_{\pm 0.03}$ & $1.22_{\pm 0.03}$ & $1.40_{\pm 0.03}$ & $1.25_{\pm 0.08}$ & 1.32 \\
\hline 
473-044754    &      &      &      &                   &      \\
\hline
$B-V    $     & $1.42_{\pm 0.06}$ & $1.37_{\pm 0.06}$ & $1.49_{\pm 0.06}$ & $1.43_{\pm 0.04}$ & 1.49 \\
$V-R_{c}$     & $0.90_{\pm 0.04}$ & $0.96_{\pm 0.04}$ & $0.95_{\pm 0.04}$ & $0.94_{\pm 0.02}$ & 0.93 \\
$B-R_{c}$     & $2.32_{\pm 0.05}$ & $2.33_{\pm 0.05}$ & $2.45_{\pm 0.05}$ & $2.37_{\pm 0.04}$ & 2.42 \\
\hline  

\end{tabular}
\end{table}

\begin{table}
\centering
\caption{The mean colors correction terms for Batch \#1, Batch \#2 and Batch \#3 derived from colors of Table~\ref{t02a}.}
\label{t02b}
\begin{tabular}{lccc}
\hline
     & Batch \#1 & Batch \#2 &  Batch \#3  \\
\hline
$B-V    $     & 0 & $+0.01_{\pm 0.03}$ & $-0.16_{\pm 0.07}$  \\
$V-R_{c}$     & 0 & $-0.06_{\pm 0.01}$ & $-0.05_{\pm 0.006}$ \\
$B-R_{c}$     & 0 & $-0.05_{\pm 0.03}$ & $-0.21_{\pm 0.07}$  \\
\hline
\end{tabular}
\end{table}

\begin{figure}
\includegraphics[width=1.0\hsize]{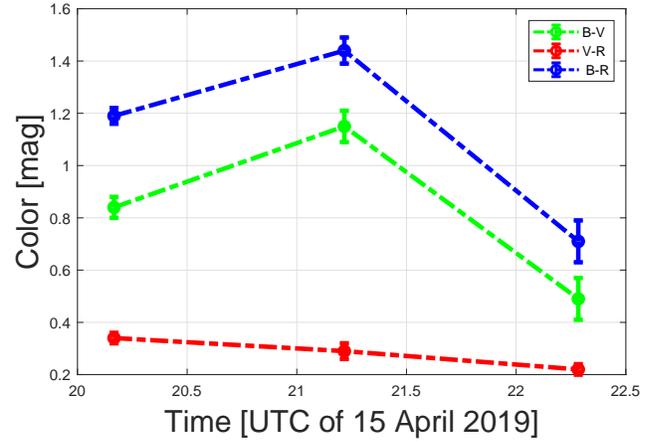}
\caption{Gault's colors variations over time.}
\label{f05c}
\end{figure}

\begin{figure}
\hspace*{-0.6cm}\includegraphics[width=1.0\hsize]{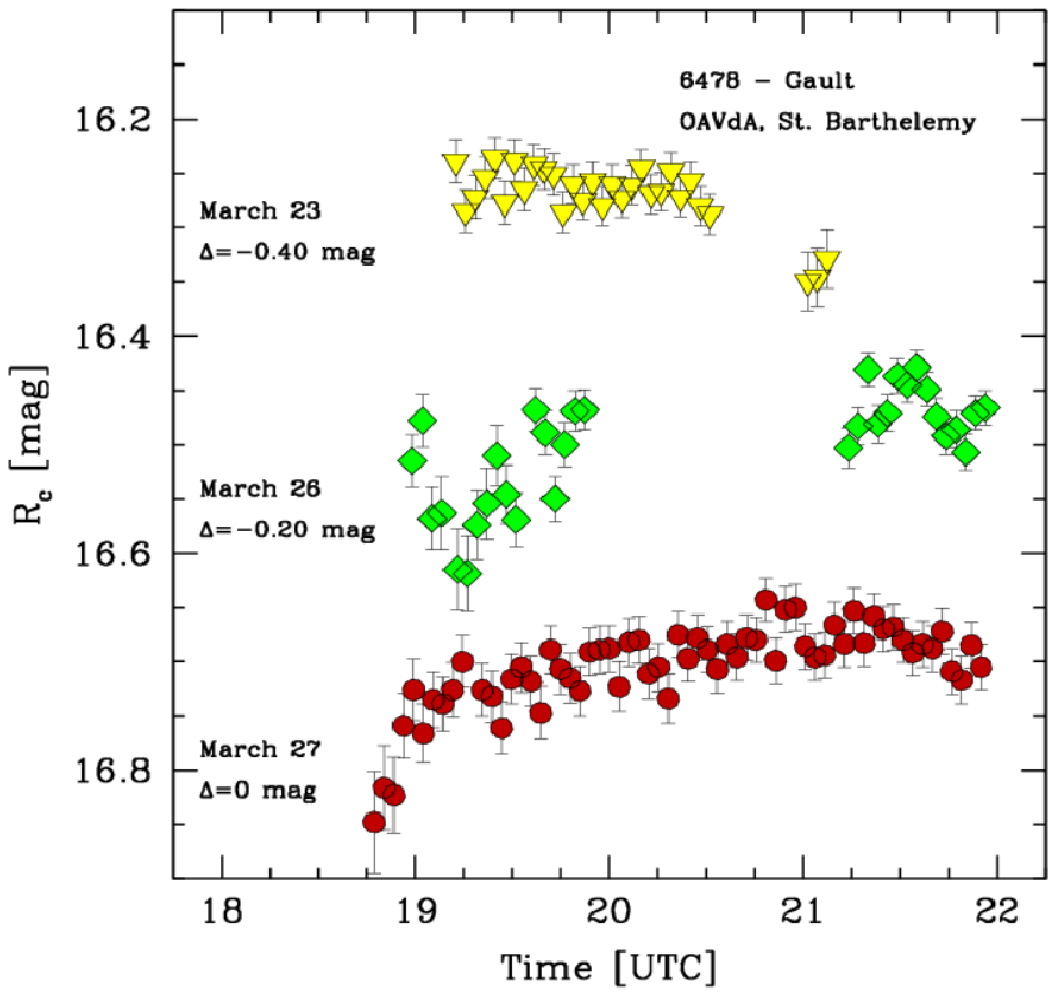}
\hspace*{-1.2cm}\includegraphics[width=1.1\hsize]{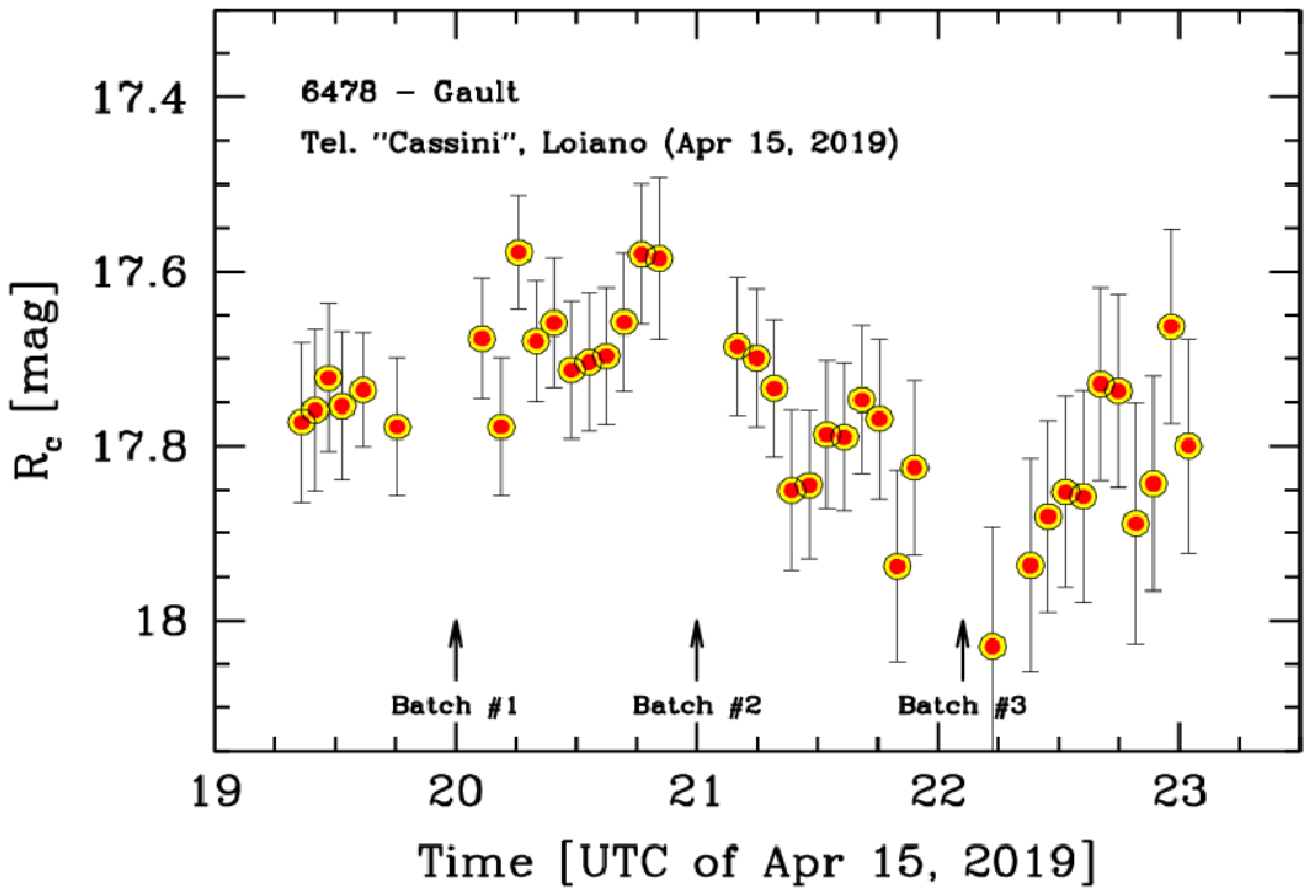}
\caption{Gault's observing sessions of 2019 March 23-27 from OAVdA (the mean uncertainty for 
the three sessions are, respectively, 0.020, 0.022 and 0.023 mag) and 2019 April 15 from Loiano (mean uncertainty 0.09 mag),
are summarized in the upper and lower panels, respectively. The $R_c$ magnitude scale is reproduced,
throughout, from the local CMC-15 calibration, according to eq.~(\ref{eq:rcmc15}).
Along the OAVdA observations, the asteroid was about its Earth oppostion, at orbital phase angle $\phi \sim 12.9^{\circ}$, 
a figure that increased to $\phi \sim 21.4^{\circ}$ for the Loiano data. The $B,V$ magnitude sampling from Batch \#1-3 
observations is marked on the plot. Note a substantial difference in lightcurve amplitude and shape between the two 
observing sessions. See text for a discussion.}
\label{f05}
\end{figure}

The apparent Gault colors along the three observing windows are displayed in Table~\ref{t03},
together with their average values. As far as the latter ones are considered, our colors
are fully consistent with \citet{Man-ToHui2019}, after correcting the latter ones to the 
Johnson-Cousins system, according to \citet{bessel1979}).
One has to report, however, the evident trend toward ``bluer color'' along the Table~\ref{t03} 
observations (see Fig.~\ref{f05c}), with the asteroid to become some $\Delta (B-V) \sim 0.35 \pm 0.09$~mag bluer toward
the minimum luminosity (see Fig.~\ref{f05}). This trend was not present in the UCAC4 stars of Table~\ref{t02a},
where, on the contrary, there is a little red-shift due to veils, as expected. We consider this bluer color a real effect: probably 
there is a bluer region on the Gault surface that we had observed during the session thanks to asteroid rotation. 
Indeed, the temporal difference between Batch \#1 and Batch \#3 is about 2 h, more than half of the best rotation period we estimated 
for Gault (see section~\ref{subsec:period}). This colors variation towards blue is consistent with what was found by 
\citet{marsset2019}, that in NIR spectroscopic observations of March 31, 2019 found Gault bluer than similar observations 
of April 8, 2019. In this last observation the Gault spectrum appears an S-type asteroid, compatible with Phocaea's spectrum.
To our knowledge, no one had observed such a marked change in Gault's color during the same session.
It is assumed as a rule  that the surface of  asteroids is uniform so finding these changes, as for NEA (297274) 1996 SK \citep{lin2014},
is very interesting.

\begin{table}
\centering
\caption{Gault's $BVR_{c}$ colors along the observing night of 2019 April 15, and comparison with \citet{Man-ToHui2019}. 
The bluer color from Batch \#1 to Batch \#3 is evident.}
\label{t03}
\begin{tabular}{lccccc}
\hline
 & Batch $\#1^{a}$ & Batch $\#2^{a}$ &  Batch $\#3^{a}$ & Average & MT19$^{b}$ \\
\hline
$B-V    $     & $0.84_{\pm 0.04}$ & $1.15_{\pm 0.06}$ & $0.49_{\pm 0.08}$ & $0.82_{\pm 0.3}$ & $0.79_{\pm 0.06}$ \\
$V-R_{c}$     & $0.34_{\pm 0.02}$ & $0.29_{\pm 0.03}$ & $0.22_{\pm 0.02}$ & $0.28_{\pm 0.06}$ & $0.31_{\pm 0.02}$ \\
$B-R_{c}$     & $1.19_{\pm 0.03}$ & $1.44_{\pm 0.05}$ & $0.71_{\pm 0.08}$ & $1.11_{\pm 0.4}$ & $1.10_{\pm 0.06}$ \\
\hline
\multicolumn{6}{l}{$^{a}$ After cloud veils correction, as discussed in the text}\\ 
\multicolumn{6}{l}{$^{b}$ As from \citet{Man-ToHui2019}}\\ 
\end{tabular}
\end{table}

\begin{figure}
\hspace*{-0.8cm}\includegraphics[width=1.0\hsize]{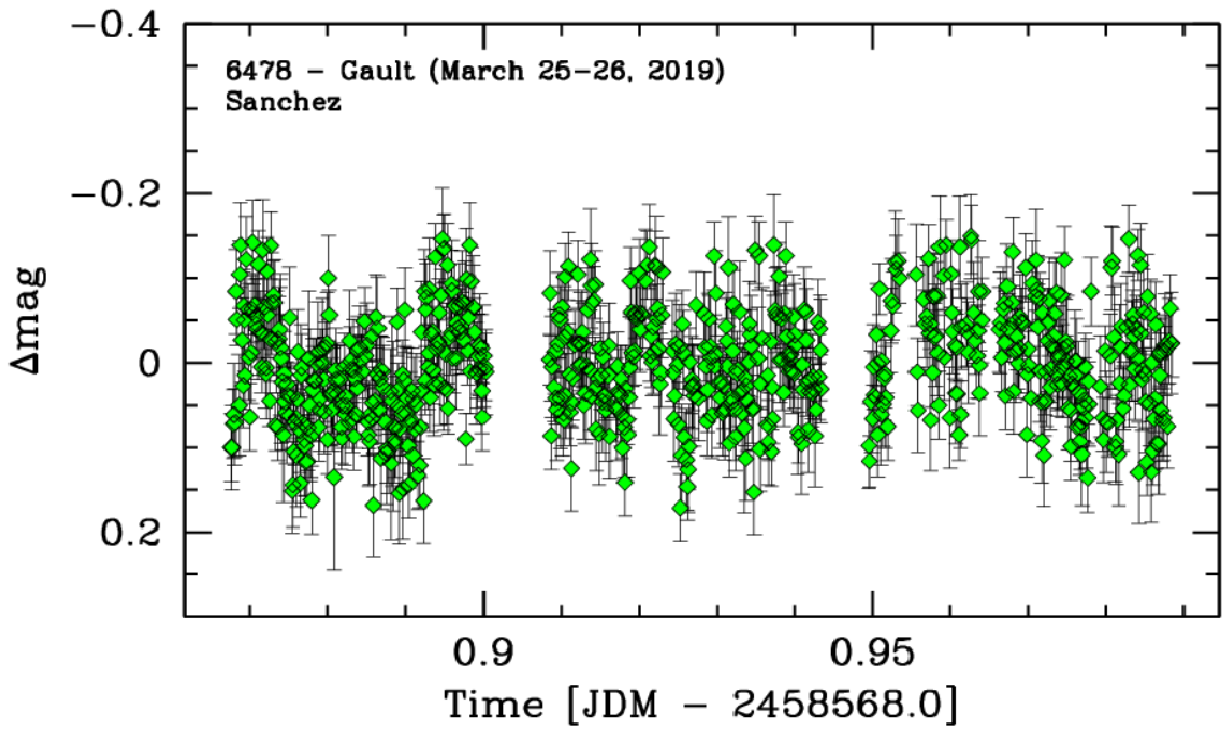}
\hspace*{-0.8cm}\includegraphics[width=1.0\hsize]{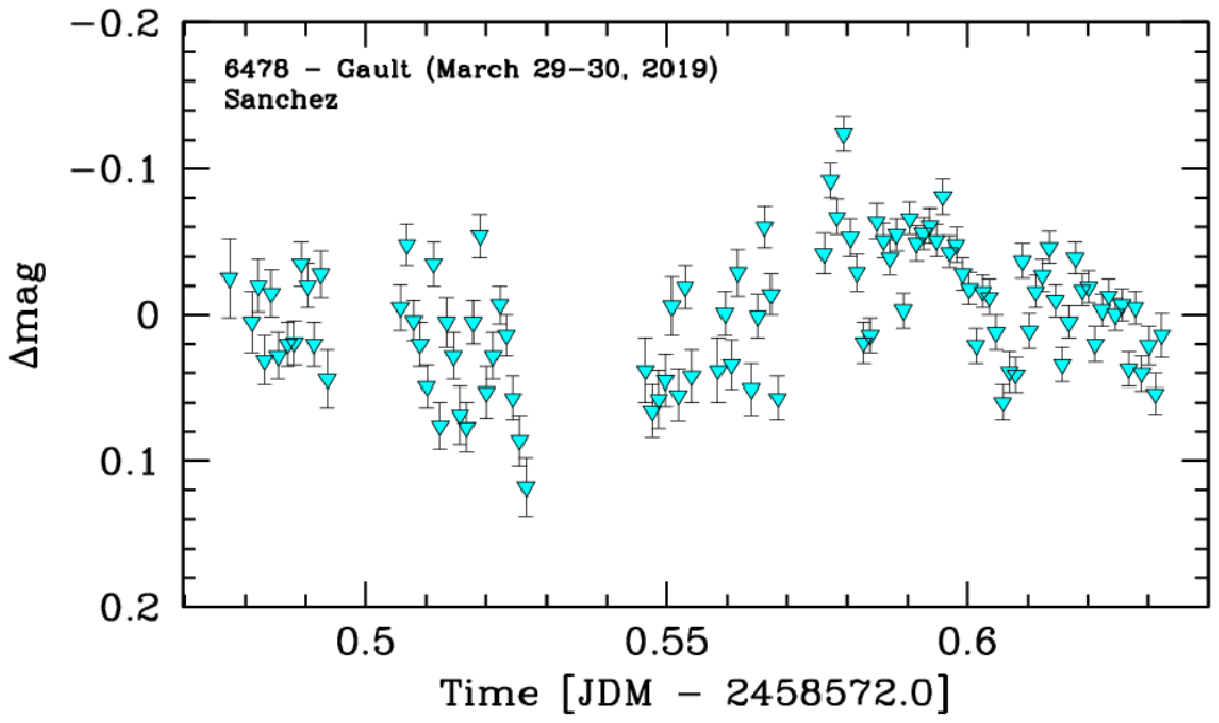}
\caption{Gault's observing sessions of 2019 March 26 and 30 from \citet{sanchez2019} are 
summarized in the upper and lower panels, respectively.}
\label{f05b}
\end{figure}

\section{Derived lightcurve}
\label{sec:lightcurves}

A general summary of the OAVdA and Loiano observations is summarized in the two panels of Fig.~\ref{f05}.
As far as the OAVdA data set is concerned, a first outstanding feature of Gault's observed 
lightcurve along all the three nights of 2019 March 23, 26 and 27 is a quite regular trend
with the object almost steady at a ``flat'' maximum interspersed with ``spiky'' minima, 
where magnitude gets some 0.1-0.15~mag fainter.

This feature, strongly reminiscent of the photometric behaviour of eclipsing binary stars, 
closely recalls a similar trend seen weeks before by the Indian 
HCT and ESA OGS telescopes, as reported by \citet[][see their Fig. 3]{kleyna2019}. A change
of status occurs, however, in the April observations from Loiano (lower panel of the figure) where, on the
contrary, the asteroid variation shows a smoother ``sinusoidal'' lightcurve and  
much larger amplitude (i.e. $A_{Rc} \sim 0.5$~mag).

Such a strong change in the lightcurve amplitude and shape prevented us from put all observing 
sessions together in a coherent period analysis. However, one may argue that this photometric 
behaviour is typical of an elongated body as the orbital phase angle ($\phi$) increases. 
In fact, the observations from the HCT and OGS telescopes, and from OAVdA as well were taken close 
to asteroid's opposition, at the mean orbital phase angle $\phi \sim 10^{\circ}\pm 3^{\circ}$ and $12.9^{\circ}$, 
respectively, while from Loiano we observed at a phase angle $\phi = 21.4^{\circ}$. 

A distinctive relatioship is recognized for asteroids of different taxonomic type between
amplitude and orbital phase angle \citep{zappala1990} in the form: 

\begin{equation}
A(\phi) = A(0)\,(1+ m\phi).
\label{eq:APR0}
\end{equation}
In the equation, $A(0)$ is the lightcurve amplitude (in mag) at the opposition (namely at $\phi = 0^o$).
If we express $\phi$ in degrees, then the scaling coefficient $m$ depends on 
the taxonomic type and can be empirically calibrated \citep{zappala1990} as $m = 0.030, 0.015$ and 0.013,
respectively for S-, C- and M-type asteroids. 

If we enter l.h. term of eq.~(\ref{eq:APR0}) with the amplitude observed from Loiano, that is 
$A(21.4^\circ)\approx 0.5$ mag, which is our best value as observed when Gault's activity was decreasing, 
then an opposition value of $A(0) = 0.35_{\pm 0.05}$ mag is inferred, accounting for the full range
of $m$ along the taxonomic class. If we assume the magnitude variation to be fully induced by
a change of reflective surface in a ``cigar-shaped'' ellipsoid (with fixed albedo), then the $(b/a)$ ratio
can be constrained as $(b/a) \approx 10^{-0.4\,A(0)} = 0.73_{\pm 0.03}$. According to Footnote~\ref{fn1}
definition, this leads to a plausible range for body's (sagittal) eccentricity of $\epsilon \approx 0.68_{\pm 0.03}$. 
This estimate implies that the lightcurve amplitude is due entirely to the asteroid shape. 
If there are albedo patches on surface, as discussed above, the elongation will be smaller.

\begin{figure}
\hspace*{-0.5cm}\includegraphics[width=1.0\hsize]{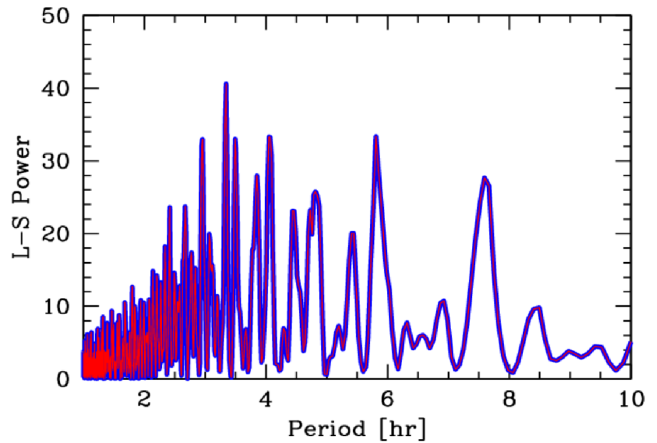}
\caption{The Lomb-Scargle periodogram of spectral power versus period (between 0.5 and 10 hours) 
for OAVdA's sessions and Sanchez 30 March. The best period is $3.34 \pm 0.02$ h.}
\label{f06c}
\end{figure}

\subsection{Rotation and spin-barrier critical period}
\label{subsec:period}

\begin{figure}
\hspace*{-0.8cm}\includegraphics[width=1.1\hsize]{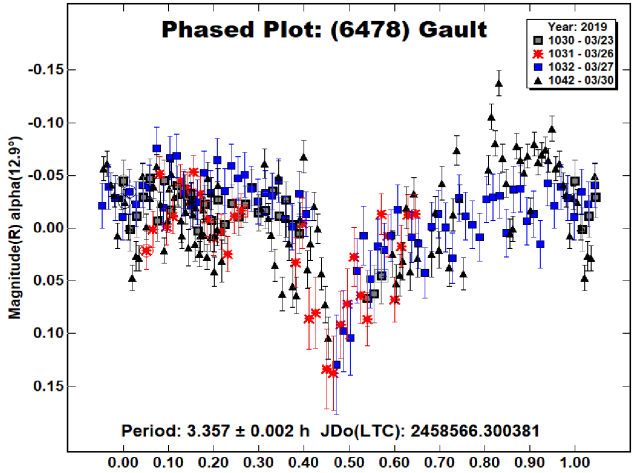}
\caption{The phased lightcurves of the March 23, 26, 27 OAVdA and March 30 Sanchez according to the
Falc algorithm implemented in MPO Canopus. The best period is 3.36 h very near to the $3.34 \pm 0.02$ h of Fig.~\ref{f06c}.}
\label{f07}
\end{figure}

Given our sparse data set, a more pondered statistical approach was pursued to constrain the possible periodicity 
in Gault's lightcurve. The OAVdA data are better suited for this excercise for they span a larger
timeline, between March 23 and 27. On the contrary, the Loiano observations
only restrain to a 4~hours interval, although they more firmly appear to constrain to allowed range of 
possible period values, yet hardly shorter than 3 hours (see Fig.~\ref{f05}).
To further extend the temporal coverage of the OAVdA's photometry, data taken from \citet{sanchez2019} on 
March 26 and 30, 2019 were also used (see Fig.~\ref{f05b}) in our analysis. We did a Lomb-Scargle analysis, 
between 0.5 and 10 h, of the Gault light curve using all the OAVdA and Sanchez sessions. 
If all sessions are used, the dominant period is 1.3 h completely given by the March 26 Sanchez session.  
This peak is a fake, due to two interruptions in the observations at a distance of 0.05 days, or about 1.2 h which gives a false periodicity to a substantially flat light curve. Another problem with the March 26 Sanchez session is that the average error on the mag is 0.05, while for the OAVdA and Sanchez sessions on March 30 it goes from 0.015 to 0.023. So Sanchez's March 26th session is twice as noisy as the others and it makes sense to 
remove it from the analysis because it is not comparable to the others.
Removing this session the best period is $3.34 \pm 0.02$ h (see Fig.~\ref{f06c}). In the periodogram remains a widened peak around 7-7.5 hours, which could correspond to the period of a hypothetical binary system, see section~\ref{subsec:binay_system} for a more detailed discussion.
This value also confirms the \citet{ferrin2019} preliminary estimate from his own photometry, and the 
period is also compatible with the Loiano observations, as evident from the plot of Fig.~\ref{f05}. 
Indeed, a Lomb-Scargle analysis of the Loiano dataset also shown a peak around a period of about 3.4 hours. 
Our results were also corroborated by independently cross-checking the OAVdA and Sanchez data with the
{\sc Falc} Fourier analysis algorithm by \citet{harris1989}, implemented in the MPO {\sc Canopus} 
package. The resulting MPO Canopus lightcurves from OAVdA and Sanchez are shown in Fig.~\ref{f07}. 
The best period is 3.36 h very close to the 3.34 h period that we had found using Lomb-Scargle. 
One major concern deals with the lack of any evident ``secondary'' minimum, about midway from two ``primary'' 
minima (i.e. ``double-peaked'' lightcurve), as usual for an asteroids. Probably Gault's reflectance 
have been heavily affected by dust activity which may have partially erased the lightcurve characteristics.
In a disrupting ``rubble-pile'' structure model, a glance to Fig.~\ref{f02} clearly points to an asteroid bulk density 
$\rho \la 1.2 ~\textrm{g}~\textrm{cm}^{-3}$, a value compatible with a internally fragmented S-type asteroid, 
i.e. with large macroporosity. As a main conclusion, our analysis definitely rules out the spin-barrier 
classical value of about 2 hours, as claimed by \citet{kleyna2019}.

\begin{figure}
\hspace*{-0.8cm}\includegraphics[width=1.1\hsize]{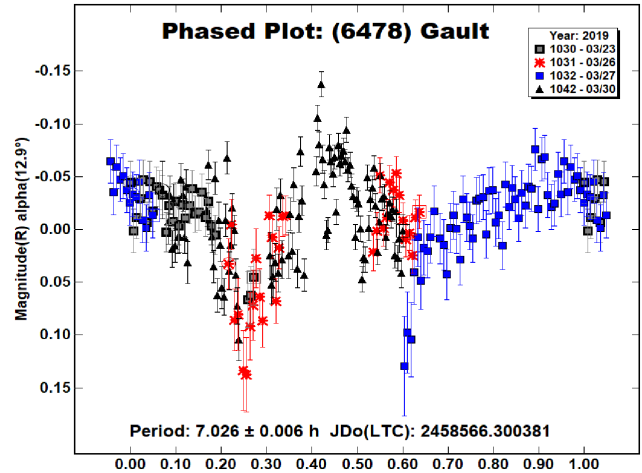}
\caption{Like Fig.~\ref{f07}, according to the $\sim 2\times$ best period values. This phased lightcurve is compatible with
a contact binary system with equal components \citep{descamps2008}.}
\label{f08}
\end{figure}

\subsection{A merging binary system?}
\label{subsec:binay_system}

Patching absorption by Gault's surrounding dust layers could naturally give reason of the 
the lack of any ``secondary'' minimum in the phased lightcurves of Fig.~\ref{f07}\footnote{Actually, 
in a dust-free ``cigar-shaped'' ellipsoid of fixed albedo, spinning around the principal momentum axis,
one {\it must} expect ``secondary'' minimum to be of equal amplitude than the ``primary'' one, 
both being generated by the opposite end-to-end extrema of the spinning ``cigar''.} and the so
erratic luminosity trend discussed in the \citet{kleyna2019} paper, as well.
Alternatively, we can match the expected ``double-peaked'' photometric trend by moving on 
the $\sim 2\times$ period pattern, with a period of about 7 hours. The resulting phased lightcurve 
of the OAVdA and Sanchez data with MPO Canopus, is shown in Fig.~\ref{f08}. Such new physical scenario could explain 
Gault's activity in terms of a near-contact binary that merge itself in a contact binary through 
the loss of angular momentum due to BYORP effect \citep{quanzhi2019}. 
Indeed a careful analysis of Fig.~\ref{f08} may recall a contact binary system of two elongated bodies of similar size
whose orbital plane is tilted enough with respect to our point of view such as to avoid 
full occultation between the two components \citep[see e.g.][for illustrative examples]{descamps2008}.
Note that the second minimum in Fig.~\ref{f08} does not fall exactly at the 0.75 phase as expected for a contact binary system, 
probably the lightcurve is ``dirty'' as a result of Gault's activity (with OAVdA's session only the second minimum fall in 0.75 phase).
In this case, if we assume the same bulk density ($\rho_G$) and size ($R_G$) for the two Gault's components,
orbiting at a distance $n\,R_G$ apart,\footnote{In our notation we have a contact binary if $n = 2$, that
is if the two asteroid components are separated by twice their reference radius $R_G$.}
then the Kepler third law provides:

\begin{figure}
\includegraphics[width=\hsize, clip=]{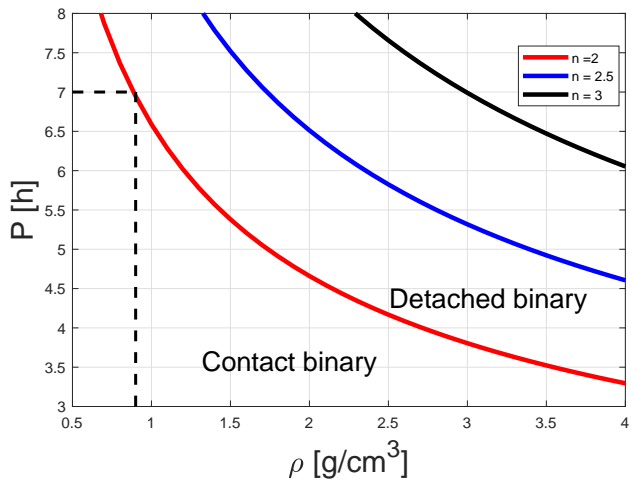}
\caption{The expected $P$ vs. $\rho$ relationship for a close binary system with the asteroid
consisting of two components of similar size and mass, according to eq.~(\ref{eq:binary}).
The component distance is parameterized in terms of multiple ``$n$'' of the body's reference radius, 
$R_G$, as in eq.~(\ref{eq:bin}). Accordingly, a contact system is obtained for $n=2$, while
for $n=3$ the two asteroid components are orbiting one $R_G$ apart. The nominal periodicity
of case P=7~h is singled out in the plot, with an implied density
for Gault of $\rho \sim 1.0$~g\ cm$^{-3}$, in case of a contact binary system.
}
\label{f09}
\end{figure}

\begin{equation}
\frac{4\,\pi^2}{P^2} = \rho_G\,G\frac{(8/3)\,\pi R_G^3}{(n\,R_G)^3},
\label{eq:bin}
\end{equation}
or
\begin{equation}
P = \sqrt{\frac{3\pi\, n^3}{2\,G}\frac{1}{\rho}} \sim \frac{2.33\,n^{3/2}}{\sqrt{\rho}} \quad {\rm [hr]}
\label{eq:binary}
\end{equation}

Figure~\ref{f09} summarizes our results for the full range of possible configurations.
In case of a preferred fiducial period of $P = 6.7$~h or larger, a contact double asteroid could be
admitted with an implied bulk density $\rho \la 1.0$~g\ cm$^{-3}$, as marked in the figure.
A much larger value for $\rho$ would however allowed in case of a close but semi-detached system.


\section{Summary and conclusions}

In this paper we comprehensively reviewed the observations made in early 2019 on the new active asteroid 
(6478) Gault. The most likely cause is that the asteroid activity was due to reconfigurations 
after YORP spin-up. However, also binary-system merging could be invoked as the main responsible of 
Gault's outbursts. For this reason an accurate estimate of the inherent photometric periodicity could 
actually discriminate between the different scenarios. Until very recently, in their 2019 observations,
\citet{kleyna2019} proposed a spinning value about 2 hours, which implied a density of 
some 2.7~g\ cm$^{-3}$, as for a typical S-type asteroid (see Fig.~\ref{f02}). This result 
was consistent with Gault's asteroid family: Phocaea.\footnote{This classification
comes, for instance, from the AstDys-2's proper elements of Gault, see: 
{\sl https://newton.spacedys.com/astdys/-index.php?pc=1.1.6\&n=6478}.\label{ft6}}
Two NIR spectra taken by \citet{marsset2019} show deep absorption band near 1 and 2 $\mu$m consistent with an S-type asteroid,
this support the link between Phocaea collisional family and Gault. 
To better clarify the situation about the rotation period, we added fresh photometric observations 
from OAVdA, in the second half of March 2019 (see Fig.~\ref{f05}).
To extend the temporal coverage of the OAVdA's photometry, data taken from \citet{sanchez2019} on 
2019 March 26 and 30, were also used (see Fig.~\ref{f05b}). Finally we did a Lomb-Scargle analysis, 
between 0.5 and 10 h, of the Gault lightcurves using OAVdA and Sanchez sessions.
From the periodogram (see Fig.~\ref{f06c}), a best period is 
identified, namely $3.34_{\pm 0.02}$ hours, with no evident sign of any $\sim 2$~h periodicity. 
The 3.34~h period also confirms the \citet{ferrin2019} preliminary estimate and it may be taken as 
the most probable, although the other near values cannot be firmly excluded at the current state of observations. 
If this is the real context, then by invoking the spin-barrier limit, Fig.~\ref{f02} shows that Gault's bulk density should not
exceed $\rho \sim 1.2$~g\ cm$^{-3}$, compatible with a fragmented S-type asteroid.
By forcing twice a photometric period in order to fit with a ``double-peaked'' lightcurve (Fig.~\ref{f08}), 
we challenged the possibility for Gault to be a merging contact (or semi-detached) binary system 
consisting in fact of similar twin bodies. A realistic solution in this case points to a period of about 
7 hours, leading to quite a ``fluffy'' bulk density $\rho \la 1.0 ~\textrm{g}~\textrm{cm}^{-3}$, in force of eq.~(\ref{eq:binary}). 
 
The mid-April lightcurve from Loiano, sampling Gault's more quiescent status compared to March 
(see Fig.~\ref{f03} and compare with Fig.~\ref{f01}), shows a greater amplitude and a more sinusoidal 
shape compared to the OAVdA observations. Also this dataset shown a peak around a period of about 3.4 hours.
In case of constant albedo, this may be suggestive of an elongated (roughly cigar-like) shape for the body, 
with an implied (sagittal) eccentricity $\epsilon \approx 0.68_{\pm 0.03}$.

Gault colors were also assessed along the Loiano observing run, leading to the
average figures summarized in Table~\ref{t03}, namely $(B-V)=0.82_{\pm 0.3}$, $(V-R_c)=0.28_{\pm 0.06}$ 
and $(B-R_c)=1.11_{\pm 0.4}$, in quite a good agreement with \citet{Man-ToHui2019}
but with a remarkable bluer color, with the asteroid becoming much bluer toward 
the minimum lightcurve luminosity (see Fig.~\ref{f05}) of about $\Delta (B-V) \sim 0.35 \pm 0.09$~mag. 
This strange behavior is supported by the aforementioned spectroscopic observations made 
on March 31 and April 8 by \citet{marsset2019}. The first spectrum was bluer than the second one and this 
indicates a macroscopic difference of albedo in different Gault's areas. It is possible that this difference 
is due to an active area that has exposed new fresh material not been reddened by solar radiation. 
Further photometric and spectroscopic observations are needed to fully characterize this very interesting minor body.

\section*{acknowledgements}
The authors wish to thank Sanchez J.~A. for granting the use of Gault photometric data and the Astronomical Observatory of the Autonomous Region of the Aosta Valley (OAVdA), managed by the Fondazione Cl\'ement Fillietroz-ONLUS, for granting the use of the Main Telescope. Many thanks to the referee for the useful suggestions that have greatly improved the quality of the manuscript.


\bsp
\label{lastpage}
\end{document}